\documentclass[twocolumn,preprint]{elsart3p}

\usepackage{epsfig}

\begin{document}

\begin{frontmatter}
{\hfill \bf Accepted for publication in Phys. Lett. B}


\title{Critical behavior in the variation of GDR width at low temperature}


\author[label1]{Deepak Pandit},
\author[label1]{S. Mukhopadhyay},
\author[label1]{Surajit Pal},
\author[label3]{A. De},
\author {and}
\author[label1]{S. R. Banerjee\corauthref{cor}}
\corauth[cor]{Corresponding author.}
\ead{srb@vecc.gov.in}

\address[label1]{Variable Energy Cyclotron Centre, 1/AF-Bidhannagar, Kolkata-700064, India}
\address[label3]{Department of Physics, Raniganj Girls' College, Raniganj - 713358, India}

\begin{abstract}
We present the first experimental giant dipole resonance (GDR) width systematics, in the temperature region 0.8 $\sim$ 1.2 MeV for $^{201}$Tl,
a near Pb nucleus, to investigate the evolution of the GDR width in shell effect $\&$ pairing  dominated region. The extracted GDR widths are well below the predictions of shell effect corrected thermal shape fluctuation model (TSFM) and thermal pairing included phonon damping model. 
A similar behavior of the GDR width is also observed for $^{63}$Cu measured in the present work and $^{119}$Sb, measured earlier. This discrepancy is attributed to the GDR induced quadrupole moment leading to a critical point in the increase of the GDR width with temperature. We incorporate this novel idea in the phenomenological description based on the TSFM for a better understanding of the GDR width systematics for the entire range of mass, spin and temperature.

\end{abstract}

\end{frontmatter}


In the analysis of collective motion in nuclei, the study of vibrational modes presents a 
field of great scope that involves a diversity of issues concerning the structure of 
quantal many-body systems. These correspond partly to the shape oscillations of different 
multipole order and partly to the fluctuations in which 
the neutrons move collectively with respect to the protons. The prime example of the polarization 
mode of collective nuclear vibration is the giant dipole resonance (GDR) in which protons 
and neutrons oscillate out of phase \cite{hara01, gaar92}.
The phenomenon couples directly to the nuclear shape degrees of freedom and the investigation 
of its strength distribution gives a direct access to the nuclear deformations \cite{gaar92}.
Moreover, it also occurs on a time scale that 
is sufficiently small and thus has been utilized by a 
variety of experiments to study hyper-deformation in alpha cluster nuclei \cite{Dipu1}, 
Jacobi shape transition \cite{Dipu1, Maj04}, 
loss of collectivity at high temperature \cite{Bor91}, fission timescale \cite{Pet94}, 
motional narrowing \cite{Orm96} and pre-equilibrium 
giant dipole vibration \cite{kel99}. However, the evolution of the intrinsic GDR 
properties at low temperature (T$<$1.5 MeV), 
in particular its width, which corresponds to the damping of this collective vibration, 
is yet not fully understood and remains an intriguing topic.

The GDR width is of particular interest since it gives an idea about the nuclear 
shear viscosity \cite{Dang11}. 
A wealth of GDR data, built on excited states, show that the GDR width increases 
with increasing T.  The behavior
can be described reasonably well within the 
thermal shape fluctuation model (TSFM) \cite{Orm96, Kus98} for T$>$1.5 MeV. 
The model proposes that at high excitation energies the nucleus undergoes shape 
fluctuations and the GDR vibrations probe its
different shapes. As a result, under adiabatic assumption, the GDR width is then a weighted 
average of all the frequencies associated with different shapes giving rise to an 
overall broadening of the GDR width. 
Unfortunately, this widely accepted model fails to explain the experimental data 
for T$<$1.5 MeV in different mass regions \cite{Orm96, Kus98, supm11, heck03, cam03, dipu2}.
We remark here that an essential point that has been overlooked in the above 
formalism is that the GDR vibration itself 
induces a quadrupole moment causing the nuclear shape to fluctuate \cite{Sim09} even at $T$=0 MeV.
Therefore, when the giant dipole vibration having its own intrinsic fluctuation is used as a probe
to view the thermal shape fluctuations, it is unlikely to feel the thermal 
fluctuations that are smaller than its own intrinsic fluctuation. 
If this assumption is true, the experimental GDR widths should remain constant at the 
ground state values until a critical temperature ($T_c$) and the effect of the thermal
 fluctuations on the experimental GDR width 
(i.e. increase of the apparent GDR width) should appear only when it becomes greater than the 
intrinsic GDR fluctuation.  
An indication of such a behavior was seen experimentally in our recent work 
on $^{119}$Sb \cite{supm11} where the GDR width was found to be 
constant at ground state value till T$\sim$1 MeV and increased subsequently thereafter 
in complete contrast to the TSFM
which predicts gradual increase of the GDR width from the ground state value. 
Earlier, the GDR width measured in coincidence with the delayed isomer for 
$^{114}$Sn was found to be much larger than the ground state value at T$\sim$0.95 MeV
\cite{Sto89}. However, the recent GDR width measurements at even lower temperature 
in different mass region are found to be close to the ground state values 
\cite{heck03, cam03, dipu2}.
On the other hand, the TSFM attributes this suppression in $^{208}$Pb due to strong 
shell effects at low temperature, which first have to be 
dissolved before the width increases with T \cite{Orm96, Kus98}. However, no data 
exists below T=1.3 MeV in this mass region to substantiate this 
prediction where shell effects are expected to be large. In order to address the above 
issues, experimental data over a range of temperature for several nuclei is needed.

In this Letter, we report on the experimental measurement of the high-energy GDR 
$\gamma$-rays from  $^{201}$Tl, a near Pb nucleus, and $^{63}$Cu to simultaneously 

investigate the GDR width at low temperature in high and low mass region. We show, 
for the first time, that even with the inclusion of shell effects in Pb region, the 
TSFM fails to explain the data at low temperature. It can be clearly seen
that even for the nuclei in other mass ranges the same holds good. We attribute this 
effect to the competition of the intrinsic GDR vibrations and the thermal shape
fluctuations, which in turn give rise to this critical behavior of GDR width at
low temperatures below T$<$1.5 MeV, for all the nuclei in general.

\begin{figure}
\begin{center}
\includegraphics[height=10.0 cm, width=7.0 cm]{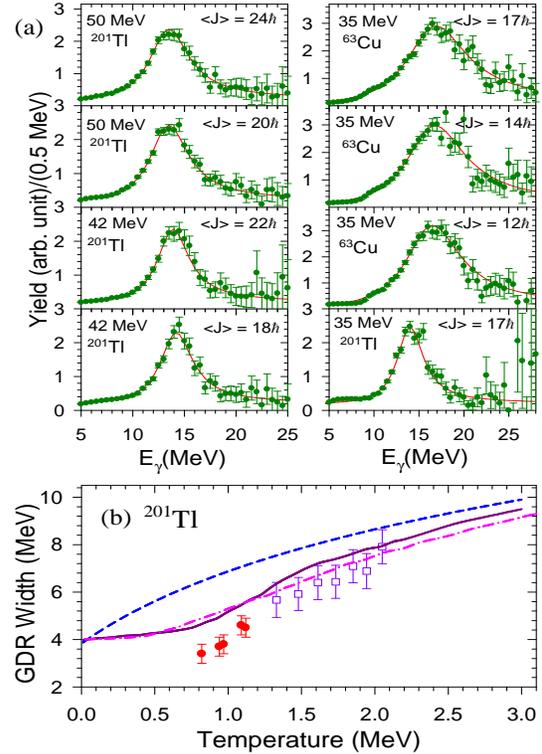}
\caption{\label{spec} (color online) (a) The linearized GDR strength functions 
(symbols) for $^{201}$Tl and $^{63}$Cu. The continuous lines are F(E$_{\gamma}$) 
used in  CASCADE prediction. (b) Temperature dependence of the GDR width for 
$^{201}$Tl. The filled circles, data from this work, are shown along with 
$^{208}$Pb\cite{Bau98} data (open squares).
The dashed, continuous and dot-dashed lines correspond to the predictions of pTSFM, 
TSFM and PDM, respectively, for $^{208}$Pb nucleus.}
\end{center}
\end{figure}

The experiments were performed at the Variable Energy Cyclotron Centre, Kolkata, using
 $^{4}$He beams produced from the K-130 room temperature cyclotron. 
Excited $^{201}$Tl and $^{63}$Cu compound nuclei were produced by bombarding 
self-supporting targets of $^{197}$Au  and $^{59}$Co, respectively. 
The initial excitation energies for $^{201}$Tl were 32.7, 39.6 and 47.5 MeV 
corresponding to incident energies of 35, 42 and 50 MeV, respectively, while it was 
38.6 MeV for $^{63}$Cu at 35 MeV incident energy. 
The critical angular momenta (L$_{cr}$) for $^{201}$Tl reactions at 35, 42 and 50 MeV 
incident energies were 16$\hbar$, 19$\hbar$ and 22$\hbar$, respectively, while it was 
14$\hbar$ for $^{63}$Cu at 35 MeV.
High-energy $\gamma$-rays from the decay of $^{201}$Tl and $^{63}$Cu were detected at 
a lab angle of 90$^\circ$ with respect to the beam axis by employing the LAMBDA 
spectrometer \cite{supm07}. The detector array, consisting of 49 BaF$_2$ detectors, 
was arranged in a 7$\times$7 matrix and kept at a distance of 50 cm from the target 
covering 2$\%$ of 4$\pi$.  
Along with the LAMBDA photon spectrometer, a 50-element BaF$_2$ 
multiplicity filter \cite{dipu3} was used to measure the discrete low energy 
multiplicity gamma rays, in coincidence with the high-energy gamma 
rays, to extract the angular momentum (J) of the compound nucleus as well as 
to get the start time trigger for the time of flight (TOF) measurement. 
The multiplicity filter was configured in two 
closed packed groups of 25 detectors each, in staggered castle type geometry, 
and placed above and below the target chamber.
The efficiency of the multiplicity setup was $\sim$ 56$\%$  
as calculated using GEANT3 \cite{brun86} simulation. 
A master trigger was generated when at least one detector each from the top and 
bottom blocks fired together in coincidence with a high-energy 
gamma ray ($>$ 4 MeV) measured in any of the large detector in the LAMBDA array. 
This ensured a selection of high-energy photons from the higher part of the 
spin distribution (Fig. \ref{spec}(a)) free from background.
The neutron-gamma discrimination of the events in the high energy spectrometer 
was achieved through the TOF measurement and the pile-up rejection was done using a 
pulse shape discrimination (PSD) technique by measuring the charge deposition over 
two integrating time intervals (30 ns and 2 $\mu$s) \cite{supm07} in each of the 
detectors. 
A VME based data acquisition system was employed to simultaneously 
record the energies and the time information of the 49 high energy detectors 
and the fold of the 50-element multiplicity filter in each event.

\begin{figure}
\begin{center}
\includegraphics[height=4.8 cm, width=7.5 cm]{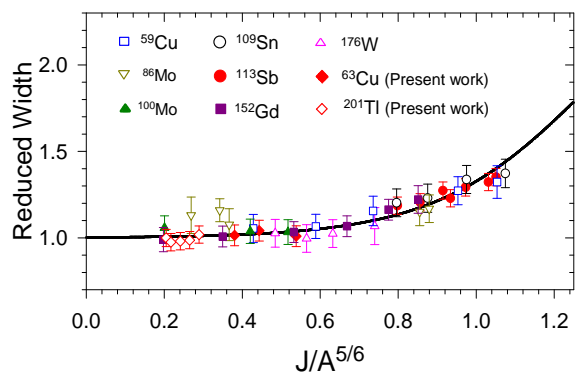}
\caption{\label{red}(color online) The reduced GDR widths are plotted against the reduced 
parameter $\xi$ = J/A$^{5/6}$ for different nuclei. The continuous line refers to scaling function  L($\xi$).} 
\end{center}
\end{figure}

The high-energy $\gamma$-ray spectra for different folds of the multiplicity 
filter were generated in offline analysis using 
a cluster summing technique \cite{supm07}. The spin distribution corresponding 
to different folds was extracted using a realistic technique based on GEANT 
simulation \cite{dipu3} and used as inputs for the statistical model calculation.
It is known that the angular momentum dependent increase of the apparent GDR width 
starts showing up above a critical spin given by the systematics 
J$_c$$\sim$0.6A$^{5/6}$ \cite{Kus98} 
and corresponds to 19$\hbar$ and 49$\hbar$ for $^{63}$Cu and $^{201}$Tl, 
respectively. Both the nuclei, $^{63}$Cu and $^{201}$Tl are populated below
this critical value in the present measurements (Fig. \ref{spec}(a)).
This is also apparent in Fig. \ref{red} where the data points lie on the flat 
part of the reduced width vs J/A$^{5/6}$ plot.  
The GDR widths were obtained from the best fit statistical model calculations 
(CASCADE \cite{cas}) along with a bremsstrahlung component folded with the 
detector response function, using a $\chi$$^2$ minimization procedure
weighted by the number of counts to take into account the exponential shape 
of the spectra. The $\chi$$^2$ minimization was carried out in the energy range 
of 8-20 MeV for $^{201}$Tl and 10-25 MeV for $^{63}$Cu. 
The bremsstrahlung component was parametrized by an exponential function 
(e$^{-E_{\gamma}/E_0}$), where the slope parameter E$_0$ was chosen according 
to the bremsstrahlung systematics \cite{nif90}. In order to emphasize 
the GDR region, the linearized GDR plots are shown in Fig. \ref{spec}(a) using 
the quantity F(E$_\gamma$)Y$^\textrm{exp}$(E$_\gamma$)/Y$^\textrm{cal}$(E$_\gamma$),
where, Y$^\textrm{exp}$(E$_\gamma$) and Y$^\textrm{cal}$(E$_\gamma$) are the 
experimental and the best fit CASCADE spectra, corresponding to the single Lorentzian 
function F(E$_\gamma$). 
The level density prescription of Reisdorf-Ignatyuk \cite{Rei81, igna75} was 
used with the asymptotic level density parameter $\widetilde{a}$=A/8 
(MeV$^{-1}$) for both the nuclei. 
Since the $\gamma$-emission from GDR decay takes place at different steps of 
the compound nuclear decay process, average values should be considered.
The average temperature of the compound nucleus associated with the GDR decay was estimated from 
$\left\langle T \right\rangle$=[($\overline{E^*}$ - $\overline{E}_{rot}$ - 
E$_{GDR}- \Delta_p$)/$a(\overline{E^*})$]$^{1/2}$, 
where $\overline{E^*}$ is the average of the excitation energy weighed over 
the daughter nuclei for the $\gamma$ emission in the GDR region. 
$\Delta_p$ is the pairing energy which is negligible at these excitation 
energies \cite{Sch99} and $\overline{E}_{rot}$ is the rotational energy computed 
at average J within the CASCADE corresponding to each multiplicity fold. 
However, at these excitation energies, the GDR decay is predominantly from the 
initial stages and the averaging only reduces the average temperature by 
$\sim$ 8$\%$. The GDR centroid energy (E$_{GDR}$) did not vary significantly with 
temperature and was centered around 13.8 and 17.0 MeV for $^{201}$Tl and $^{63}$Cu, 
respectively.

\begin{figure}
\begin{center}
\includegraphics[height=9.0 cm, width=6.5 cm]{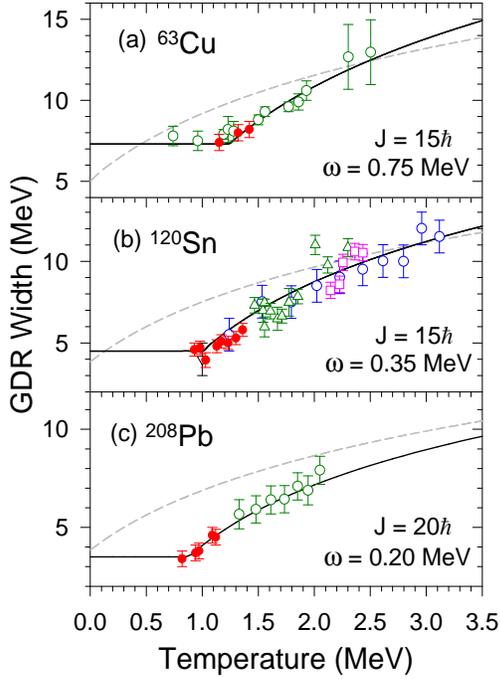}
\caption{\label{tem}(color online) The GDR widths as a function of T for
$^{63}$Cu, $^{120}$Sn and $^{208}$Pb. (a) The filled circles are the data of $^{63}$Cu 
from the present work while  open circles are from Refs\cite{Kus98, Kic87}.
(b) Our $^{119}$Sb data (filled circles) measured earlier are shown along with the data of
$^{120}$Sn (open circles\cite{Bau98}, open squares\cite{kel99}, up triangle\cite{Kus98}, 
down triangle\cite{heck03}). (c) $^{201}$Tl data (filled circles) from the present work 
along with $^{208}$Pb\cite{Bau98} data (open circles).
The dashed lines correspond to the pTSFM calculation while the continuous lines 
are the results of CTFM calculation.}        
\end{center}
\end{figure}

\begin{figure}
\begin{center}
\includegraphics[height=11.0 cm, width=7.0 cm]{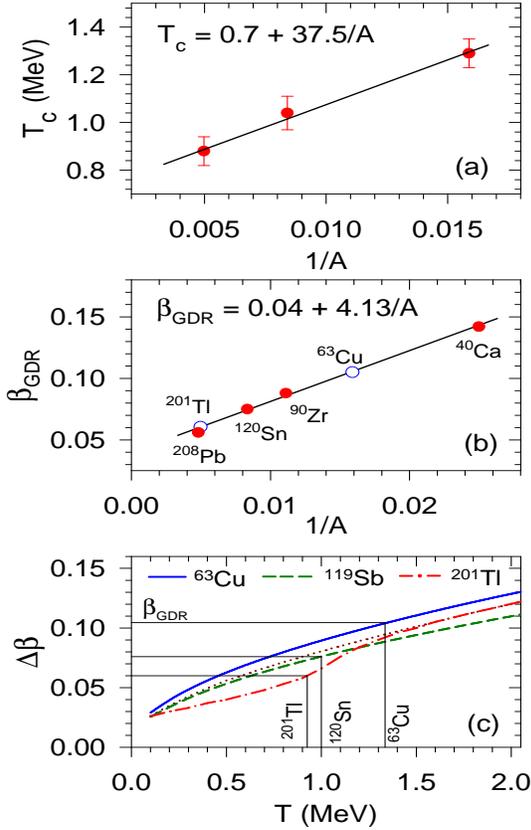}
\caption{\label{pb}(color online)(a) Critical temperature vs 1/A. Experimental data (symbols) fitted
with a linear function (continuous line).
(b) $\beta$$_{GDR}$ vs 1/A. Estimated values from the quadrupole moment (filled circles) fitted
with a linear function (continuous line). The $\beta$$_{GDR}$ values extracted from this 
systematics for $^{63}$Cu and $^{201}$Tl are shown with open circles. 
(c) The variance of the deformation as a function of T 
for $^{63}$Cu, $^{119}$Sb  and $^{201}$Tl. The dotted line represents the calculation for $^{201}$Tl without including shell effect.
The corresponding $\beta$$_{GDR}$ is compared with $\Delta$$\beta$ (thin continuous line).}
\end{center}
\end{figure}

The deduced GDR widths, 3.4, 3.7, 3.8, 4.6 and 4.5 MeV at 
$\left\langle T \right\rangle$ of 0.82, 0.94, 0.97, 1.09 and 1.12 MeV, 
respectively, from the best fit for $^{201}$Tl are shown in Fig. \ref{spec}(b) 
along with the experimental results for $^{208}$Pb \cite{Bau98}. 
The uncertainty of the extracted GDR widths was estimated to be ($\pm$ 0.45 MeV) 
by comparing the results of the statistical calculation to the measured spectra. 
This error estimate includes the statistical uncertainty, the effect of varying the 
GDR energy and the uncertainty in the non-statistical contribution.
The GDR widths 
predicted according to the phonon damping model (PDM) \cite{Dang98}, TSFM\cite{Orm96} 
and the phenomenological parametrization pTSFM\cite{Kus98} based on the TSFM 
as a function of T for $^{208}$Pb are also shown in Fig. \ref{spec}(b). 
We highlight here that the GDR widths measured in the present work provides an 
important testing ground for the theoretical models at low temperature for different 
nuclei for which data was not available earlier. 
As can be seen from Fig. \ref{spec}(b), the pTSFM (dashed line) fails completely to 
explain the experimental systematics. We emphasize that the TSFM (continuous line) 
also fails to describe the GDR width measured in the present work even after 
incorporating the shell effects. The discrepancy, therefore, clearly indicates 
that the shell effect alone cannot describe the suppression 
of the GDR width at these low temperatures and is a general feature for all the
nuclei in the entire mass range.
As a matter of fact, the microscopic PDM (dot-dashed line) as well, which
emphasizes on the importance of the coupling of the GDR  phonon to 
\textit{\textsl{pp}} and \textit{\textsl{hh}} configurations and includes 
the  effect of thermal pairing on the GDR width, cannot explain the present measurement.
Our extracted GDR widths for $^{63}$Cu, $^{201}$Tl and $^{119}$Sb \cite{supm11} 
together with $^{63}$Cu\cite{Kus98, Kic87}, $^{120}$Sn\cite{kel99, Kus98, heck03, 
Bau98} and $^{208}$Pb\cite{Bau98}, measured earlier,
are shown in Fig. \ref{tem}. 
Interestingly, it can be seen that the GDR widths for all the three nuclei 
decrease with decrease in T and reach the ground state value well above T=0 MeV, 
which prompts us to the assumption that the GDR vibration is not able 
to probe the thermal fluctuations (below $T_c$)
which are smaller than its own intrinsic fluctuation due to the GDR induced quadrupole 
moment. Only the GDR width measurement at low T in $^{114}$Sn does not follow this trend 
\cite{Sto89}. 
We extracted the critical temperatures from the experimental systematics for 
$^{63}$Cu, $^{119}$Sb and $^{201}$Tl corresponding to the ground state widths of
7.3, 4.5 and 3.5 MeV, respectively (Fig. \ref{tem}). These ground state widths
have been estimated from the recent spreading width parametrization \cite{jun08}
corresponding to the respective ground state deformations \cite{mol95}.
It is interesting to note that the critical temperature decreases with the 
increase in mass and shows a linear behavior with 1/A (Fig. \ref{pb}(a)). 

We mention here that the GDR width (3.4 MeV) measured at 0.82 MeV for $^{201}$Tl is 
systematically smaller than the ground state GDR width in $^{208}$Pb ($\sim$ 3.9 MeV) 
given by Berman and Fultz \cite{ber75} but is in good agreement with the estimated 
ground state value (3.5 MeV) considering the spreading width parametrization 
and the corresponding ground state deformation. This new spreading width parametrization \cite{jun08} 
has been obtained by separating the deformation induced widening from the spreading effect 
and requiring the integrated Lorentzian curves to fulfill the TRK sum-rule. 
It gives remarkably good match for the entire region covering the data below and 
above the particle separation energies \cite{erh10}. 
This could not be done earlier in ($\gamma$, n) reactions and the GDR widths were obtained by
just fitting the peak region of the GDR \cite{ber75}. A similar result is observed for $^{119}$Sb where the 
measured GDR widths at T $<$ 1 MeV match vary well with the estimated ground state value (4.5 MeV) 
but are  slightly smaller than the ground state value given by Berman and Fultz \cite{ber75} 
for $^{120}$Sn ($\sim$ 4.9 MeV). 
However, the GDR width measured at T$\sim$1 MeV for both $^{119}$Sb \cite{supm11}  
\& $^{120}$Sn \cite{heck03} are identical and significantly smaller than 
the ground state GDR width and require further investigation for such an unusual behavior. 
Nevertheless, the overall variation of the GDR width shows a very similar behavior with T for 
all the three nuclei (spanning almost the entire mass range) and thus one should expect a 
common phenomenon responsible for such a critical behavior in the evolution of the GDR width.

Macroscopically, the isovector GDR is interpreted as the superposition of the
Goldhaber-Teller (GT) and the Steinwedel-Jensen (SJ) modes where the former 
amounts more than the latter for all nuclei \cite{Mye87}.
In the SJ mode \cite{Ste50}, the interpenetrating and compressible neutron and 
proton fluids are constrained to 
move within a sphere with its surface effectively clamped, which does 
not affect the quadrupole moment. However, the GT mode \cite{Gol48} that assumes 
harmonic displacement of incompressible and rigid spheres of protons against 
neutrons induces a prolate shape with a 
quadrupole moment proportional to the square of the distance between the two 
spheres \cite{Sim09}. It has been shown in Ref\cite{Cam99} that
even though the equilibrium deformation of a nucleus increases with angular 
momentum, an increase of GDR width is not evident experimentally until the 
equilibrium deformation ($\beta$$_{eq}$) increases sufficiently to affect 
the thermal average. In particular, as long as  $\beta$$_{eq}$ is less than the variance 
$\Delta$$\beta$=$\left[\left\langle \beta^2\right\rangle - 
\left\langle \beta\right\rangle^2\right]^{1/2}$ 
the increase of GDR width is not significant. Similarly, the effect of thermal 
fluctuations on the experimental width should not be evident when 
$\Delta$$\beta$ due to the thermal fluctuations is smaller than the intrinsic 
GDR fluctuation ($\beta$$_{GDR}$) due to induced quadrupole moment.

The couplings between the collective vibrations such as the isovector giant dipole and
isoscalar giant quadrupole resonances have been studied in Refs\cite{Sim09, Sim03}. These couplings
are a source of anharmonicity in the multiphonon spectrum. They also affect the dipole
motion in a nucleus with static or dynamical deformation induced by a quadrupole
constraint or boost, respectively. Quadrupole moment (Q$_Q$) induced by the GDR motion
has been calculated under the framework of time dependent Hartree-Fock theory in Refs\cite{Sim09, Sim03}. 
Using the reported values for the quadrupole moments for 
$^{208}$Pb, $^{120}$Sn, $^{90}$Zr and $^{40}$Ca
as 99.0, 56.0, 46.5 and 21.4 fm$^2$, respectively,
the $\beta_{GDR}$ values were estimated considering 
$\beta$ $\propto$  Q$_Q$ /$\left\langle \overline{r}^2 \right\rangle$ 
for ellipsoidal shapes in general, where 
$\left\langle \overline{r}^L \right\rangle$=3R$^L$/(L+3). 
The estimated values are shown in Fig. \ref{pb}(b) (filled circles). 
It is interesting to note that the $\beta_{GDR}$
also decreases with increase in mass and shows a linear behavior with 1/A similar to the
critical temperature measured in the present work. However, according to our assumption,
the critical temperature should depend on the competition between $\beta_{GDR}$
and $\Delta\beta$. Hence, the variance of the deformation 
($\Delta\beta$) for $^{63}$Cu, $^{119}$Sb and $^{201}$Tl were calculated using the
Boltzmann probability $e^{-F(\beta, \gamma)/T}$ with the volume element 
$\beta^{4} sin(3\gamma) d\beta d\gamma$, according to the formalism described in
Ref\cite{Dipu1}. Interestingly, it can be seen that $\Delta\beta$ for
$^{119}$Sb and $\beta_{GDR}$ for $^{120}$Sn are about the same at T=1 MeV and matches well
with the extracted critical temperature (Fig. \ref{pb}(a),(c)). Next, the $\beta_{GDR}$
values were estimated for $^{63}$Cu and $^{201}$Tl (open circles) from the systematics in
Fig. \ref{pb}(b) and compared with the corresponding $\Delta\beta$ in Fig. \ref{pb}(c).
Most importantly, in these cases also, the temperatures at which $\beta_{GDR}$ is equal
to $\Delta\beta$ correspond to the experimentally measured critical temperatures Fig. \ref{pb}(c).
The $\Delta\beta$ values for $^{201}$Tl  were calculated for the two cases,
i.e. with and without shell effects (represented by the dot-dashed and dotted lines,
respectively, in Fig. \ref{pb}(c)). It can be clearly seen that the shell effect indeed
plays an important role in this case in correctly reproducing the experimentally measured
critical temperature. Without the inclusion of the shell effect, the values of
$\Delta\beta$ and $\beta_{GDR}$ are equal at T$\sim$0.55 MeV, whereas the experimental
result shows T$_c$$\sim$0.9 MeV.
The inclusion of the shell effect in $\Delta\beta$ for thermal
fluctuations leads to a higher T$_c$, because for temperatures $T<T_c$, $\beta_{GDR}$
dominates and only after T$_c$ the thermal fluctuations take over. Thus, the competition
of $\beta_{GDR}$ and $\Delta\beta$ giving rise to a T$_c$ and that the experimental
GDR width stays at its ground state value below T$_c$, clearly indicate that the
GDR vibration is not able to probe the thermal fluctuations that are smaller than
its own intrinsic fluctuations due to induced quadrupole moment.
It has also been shown in Ref\cite{Sim09} that the matrix element for the residual interaction
between dipole and quadrupole vibration decreases with increase in mass number
and shows a linear behavior with 1/A for Sn isotopes.
Interestingly, a similar behavior is observed in the present work where the
critical temperature shows a 1/A dependence. The appearance of a critical temperature
in the variation of GDR width could perhaps be the experimental signature
of the GDR-GQR coupling at finite T. Alternatively, in order to probe this effect
experimentally, one needs to examine the coupling of the 1$^-$ GDR to 2$^+$ states
by measuring the decay branch of GDR to the 2$+$ states at zero temperature.
However, it will be an even more difficult experimental task to identify the
GDR-GQR coupling at finite T in the statistical ensemble of states in the continuum.

Here, we also make an attempt to implement the idea of this critical behavior by modifying the phenomenological 
parametrization pTSFM \cite{Kus98}  and term it as Critical Temperature included Fluctuation Model (CTFM). 
The T dependence of the GDR width in pTSFM was parametrized as $\Gamma(T, J=0, A)$ = c($A$)$\cdot$$\ln(1+T/T_0)$ + $\Gamma_0(A)$
where c($A$)=(6.45 - A/100) and T$_0$=1 MeV. The value of $\Gamma$$_0$ is usually extracted from the measured ground state GDR width. This simple parametrization fails to represent the experimental data, particularly at low temperature, in $^{63}$Cu, $^{120}$Sn and $^{208}$Pb 
even when substantially lower values for $\Gamma_0$ were used compared to its ground state values
($\Gamma$$_0$=5 MeV for $^{63}$Cu and 3.8 MeV for $^{120}$Sn). Moreover, the model predicts gradual increase of the GDR width at all finite temperature starting from the ground state values
(dashed lines in Fig.\ref{tem}) where as the data present a completely different picture. 
However, we mention that the parametrization of Ref\cite{Kus98} was deduced at a time when there was no experimental data for GDR widths at low temperatures (T$\sim$1 MeV).
Hence, the parametrization may now be revisited and we propose
the T dependence of GDR widths in CTFM  by including the GDR induced fluctuation as 
\begin{eqnarray}
\Gamma(T, J=0, A) =  \Gamma_0(A)  \;\;\;\;\;\; T \leq T_c \nonumber
\end{eqnarray}
\begin{equation}
\Gamma(T, J=0, A) = c\left(A\right) \ln\left(\frac{T}{T_c}\right) +  \Gamma_0(A) \;\;\;\;\;\; T > T_c \label{eqn2}
\end{equation}
where, 
\begin{eqnarray}
T_c = 0.7 + 37.5/A \nonumber
\end{eqnarray}
\begin{eqnarray}
c(A) = 8.45 - A/50. \nonumber
\end{eqnarray}
Contrary to Ref\cite{Kus98}, where $\Gamma$$_0$ was based on physical choice, we propose that the apparent ground state GDR width be calculated using the ground state deformation and spreading width parametrization 
$\Gamma$$_s$=0.05E$^{1.6}_{GDR}$ \cite{jun08}  for each Lorentzian
since experimentally one could only probe the total width of the GDR (not the spherical width).
The $\Gamma$$_0$ parameters for $^{63}$Cu, $^{119}$Sb and $^{201}$Tl were estimated to be 7.3, 4.5 and 3.5 MeV, respectively, 
using the known ground state deformations \cite{mol95} (0.162, -0.12, -0.044 respectively). 
The estimated ground state values, as expected, are indeed in agreement with the actual measured values. 
The GDR width predicted by the CTFM as a function of temperature is shown in Fig. \ref{tem} (continuous lines).
We simultaneously examine the spin dependence of the width. The angular momentum dependence was parametrized through the reduced width at different temperature by a power law [$\Gamma$$_{exp}$(T,J,A)/ $\Gamma$(T,J=0,A)]$^{(T+3T_0)/4T_0}$ \cite{Kus98}. Since the coefficient of the power law was based on the physical choice of $\Gamma$$_0$ and T$_0$, it necessitates a modification in order to explain the J dependence of the width.  Using the available experimental systematics in high J region for the masses  A$\sim$60 to 200, we propose the J-dependence of CTFM as 
\begin{equation} 
\Gamma_{red} = \left[\frac{\Gamma_{exp}(T, J, A)}{\Gamma(T, J=0, A)}\right]^{\frac{T+3.3T_c}{7T_c}} = L\left(\xi\right) \label{eqn3}
\end{equation}
where, L($\xi$) = 1 + 1.8/[1 + e$^{(1.3 - \xi)/0.2}$] and $\xi$ = J/A$^{5/6}$. 
The $\Gamma$$_0$ value for $^{59}$Cu\cite{Dre95}, $^{86}$Mo\cite{Rat03}, $^{100}$Mo\cite{Kic92}, $^{109}$Sn\cite{Bra04}, 
$^{113}$Sb\cite{Sri08}, $^{152}$Gd\cite{Drc10} and $^{176}$W\cite{Mat95} nuclei 
were estimated from systematics as 6.7, 5.8, 7.3, 5.4, 5.2, 5.8 and 6.2 MeV, respectively. 
Interestingly, these estimated values agree very well with the actual measured ground state values. 
The reduced GDR widths  with the reduced parameter J/A$^{5/6}$ for different temperature and masses are shown in Fig. \ref{red}. 
It is evident that the CTFM gives
an excellent description of the GDR width systematics at low temperature as well as at high angular momentum
for the mass region A $\sim$ 60 to 208. Hence, as it appears, the experimental GDR widths are not suppressed
rather TSFM overpredicts the GDR width at low temperature since it does not take 
into account the intrinsic GDR fluctuation induced by the GDR quadrupole moment. 
The experimental observation of this critical behavior in almost the entire mass range and invoking the 
idea of competition between the thermal and the intrinsic GDR fluctuations in explaining the critical behavior could be, 
in an indirect way, the experimental verification for the coupling of GDR-GQR in nuclei at finite temperatures. 
However, more experimental and theoretical work needs to be done.

In summary, we present the first experimental measurement of GDR width for $^{201}$Tl, a near Pb nucleus, in the unexplored temperature 
region 0.8-1.2 MeV and find that the extracted GDR widths are well below the prediction of TSFM even after 
including the shell effects. 
Similar results are also observed for $^{63}$Cu (present measurement) and $^{119}$Sb (recent measurement).
It seems that the GDR induced quadrupole moment plays a decisive 
role as the GDR is not able to view the thermal fluctuations which are smaller than its own intrinsic 
fluctuation. When this effect is taken into account, the phenomenological CTFM gives a better description of the 
global systematics of the GDR width.



\end{document}